\documentclass[prl,showpacs,twocolumn,,superscriptaddress]{revtex4}
\usepackage{amsmath,graphicx,bm,float}
\newcommand{\ket}[1]{|{#1}\rangle}
\newcommand{\bra}[1]{\langle{#1}|}

\begin{document}
\title{Experimental Activation of Bound Entanglement}
\author{Fumihiro Kaneda}
\affiliation{Research Institute of Electrical Communication, Tohoku University, Sendai 980-8577, Japan}
\author{Ryosuke Shimizu}
\affiliation{Research Institute of Electrical Communication, Tohoku University, Sendai 980-8577, Japan}
\affiliation{Center for Frontier Science and Engineering, University of Electro-Communications, Tokyo 182-8585, Japan}
\author{Satoshi Ishizaka}
\affiliation{Graduate School of Integrated Arts and Sciences, Hiroshima University, Higashi-Hiroshima 739-8521, Japan}
\author{Yasuyoshi Mitsumori}
\affiliation{Research Institute of Electrical Communication, Tohoku University, Sendai 980-8577, Japan}
\author{Hideo Kosaka}
\affiliation{Research Institute of Electrical Communication, Tohoku University, Sendai 980-8577, Japan}
\author{Keiichi Edamatsu}
\affiliation{Research Institute of Electrical Communication, Tohoku University, Sendai 980-8577, Japan}

\begin{abstract}
Entanglement is one of the essential resources in quantum information and communication technology (QICT).
The entanglement thus far explored and applied to QICT has been pure and distillable entanglement. 
Yet there is another type of entanglement, called `bound entanglement', which is not distillable by local operations and classical communication (LOCC). 
We demonstrate the experimental `activation' of the bound entanglement held in the four-qubit Smolin state, 
unleashing its immanent entanglement in distillable form, with the help of auxiliary two-qubit entanglement and LOCC. 
We anticipate that it opens the way to a new class of QICT applications that utilize more general classes of entanglement than ever, including bound entanglement.


\end{abstract}
\maketitle


Quantum entanglement, one of the most counterintuitive effects in quantum mechanics \cite{Einstein35}, plays an essential role in quantum information and communication technology. 
Thus far, many efforts have been taken to create multipartite entanglement in photon polarization \cite{Walther2005, Lu2007, Gao2010, Wieczorek2009, Prevedel2009}, quadrature amplitude \cite{Furusawa2009}, and ions \cite{Haffner2005}, for demonstration and precise operation of quantum protocols. 
These efforts have mainly concentrated on the generation of pure entangled states, such as GHZ \cite{Greenberger1989}, W \cite{Dur2000}, and cluster \cite{Briegel2001} states. 
By contrast, bound entanglement \cite{Horodecki1998} could not be distilled into pure entangled states, and had been considered useless for quantum information protocols such as quantum teleportation \cite{Boumeester1997, Furusawa1998}. 
However, it is interesting that some bound entanglement can be distilled by certain procedures \cite{Smolin2001} or interaction with auxiliary systems \cite{Acin2004, Shor2003}. 
These properties provide new quantum communication schemes, for instance, remote information concentration \cite{Murao2001}, secure quantum key distribution \cite{Horodecki2005}, super-activation \cite{Shor2003}, and convertibility of pure entangled states \cite{Ishizaka2004}.  

Recently, a distillation protocol from the bound entangled state, so called `unlocking' \cite{Smolin2001}, has been experimentally demonstrated \cite{Amselem2009,Lavoie2010}. 
In this protocol, as depicted in Fig. 1 (a), four-party bound entanglement in the Smolin state \cite{Smolin2001} can be distilled into two parties (e.g., A and D) when the other two parties (e.g., B and C) come together and make joint measurements on their qubits. 
The unlocking protocol, in principle, can distill pure and maximal entanglement into the two qubits. However, it does not belong to the category of LOCC, since the two parties have to meet to distill the entanglement.

\begin{tiny}
\begin{figure}[t!]
   \includegraphics[width= \columnwidth, clip]{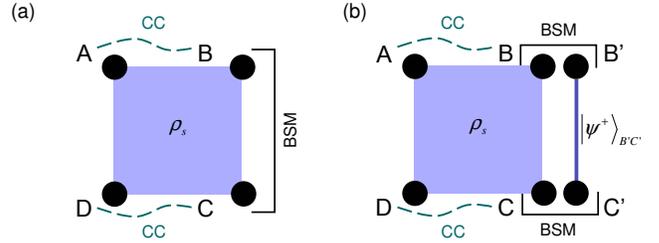} 
   \label{ }
   \caption{Principle of the distillation of the bound entanglement in the Smolin state $\rho_s$. Each circle represents a qubit, and the blue line and squares show the entanglement of parties. BSM, Bell state measurement; CC, classical channel. (a) Unlocking. (b) Activation of the bound entanglement. Both of them can distill entanglement from the Smolin state; however, the activation protocol can be carried out under LOCC, while the unlocking needs two parties coming together and making joint measurements.     }
\end{figure}
\end{tiny}


The activation of bound entanglement that we demonstrate here is another protocol by which one can distill the Smolin-state bound entanglement by means of LOCC. The principle of the activation is sketched in Fig. 1 (b). 
%
Consider four parties, A, B, C, and D each of which has a qubit in the Smolin state. 
The Smolin state is a statistically equal mixture of pairs of the four Bell states, and its density matrix $\rho_s$ is given by

\begin{align}
\rho_s &= \sum_{i=1}^{4} |\phi^i\rangle \langle \phi^i |_{AB}\otimes |\phi^i\rangle \langle \phi^i |_{CD} \notag \\ 
       &= \sum_{i=1}^{4} |\phi^i\rangle \langle \phi^i |_{AC}\otimes |\phi^i\rangle \langle \phi^i |_{BD} \notag \\ 
       &= \sum_{i=1}^{4} |\phi^i\rangle \langle \phi^i |_{AD}\otimes |\phi^i\rangle \langle \phi^i |_{BC},  
\end{align}
where $|\phi^i\rangle \in \{ |\phi^{\pm} \rangle ,|\psi^{\pm} \rangle \}$ are the two-qubit Bell states given by   
\begin{align}
|\phi^{\pm} \rangle &= \frac{1}{\sqrt{2}} \left( |00 \rangle \pm |11 \rangle \right) \notag \\ 
|\psi^{\pm} \rangle &= \frac{1}{\sqrt{2}} \left( |01 \rangle \pm |10 \rangle \right), 
\end{align}
where  $\ket{0}$ and $\ket{1}$ are the qubit bases. 
Since the $\rho_s$ state is symmetric with respect to the exchange of any two parties, $\rho_s$ is separable in any two-two bipartite cuts. 
This implies that there is no distillable entanglement in any two-two bipartite cuts: 
$D_{AB|CD}(\rho_s)=D_{AC|BD}(\rho_s)=D_{AD|BC}(\rho_s)=0$, where $D_{i|j}(\rho)$ is the distillable entanglement of $\rho$ in an $i|j$ bipartite cut. 
In addition to the Smolin state $\rho_s$, two of the parties (e.g., B and C) share distillable entanglement in the two-qubit Bell state (e.g.,  $\ket{\psi^+}_{B'C'}$). 
Hence, the initial state $\rho_I$ is given by
\begin{align}
\rho_I &= \rho_s \otimes \ket{\psi^+}\bra{\psi^+}_{B'C'} \notag \\ 
       &= \sum_{i=1}^{4} \ket{\phi^i}\bra{\phi^i}_{AD}\otimes \ket{\phi^i}\bra{\phi^i} _{BC} \otimes \ket{\psi^+}\bra{\psi^+}_{B'C'}. 
\end{align}
The state $\rho_s$ or $\ket{\psi^+}_{B'C'}$ gives no distillable entanglement into A and D: $D_{A|D}(\rho_s) = D_{A|D}(\ket{\psi^+}_{B'C'})=0$, since $D_{A|D}(\rho_s) \leq D_{AB|CD}(\rho_s)=0$. 
To distill entanglement into A and D, the Bell state measurements (BSMs),
the projection measurements into the Bell bases, 
are taken for the qubits B-B' and C-C', and the results are informed A and D via classical channels. 
Due to the property that A-D and B-C share the same Bell states $\ket{\phi^i}_{AD}\otimes\ket{\phi^i}_{BC}$ in $\rho_s$, the result of the BSMs in each $\ket{\phi^i}_{BC}\otimes \ket{\psi^+}_{B'C'}$ can tell the type of the Bell state shared by A and D. 
Table I shows the list of the resulting states shared by A and D for all the possible combinations of the result of the BSM of B-B' and C-C'.  
Given this information one can determine the state shared by A and D and then convert any $\ket{\phi^i}_{AD}$ into $\ket{\psi^-}_{AD}$ by local unitary operations. 
Hence, the activation protocol can distill entanglement from the Smolin state by four parties' LOCC with the help of the auxiliary two-qubit Bell state. 
This is in strong contrast to the unlocking protocol (see Fig. 1 (a)), which requires non-local joint BSM between the two parties (B and C). 
It is noteworthy that in our activation protocol, the distillable entanglement between A and D is superadditive:
\begin{equation}
D_{A|D}(\rho_I)  > D_{A|D}(\rho_s) +D_{A|D}(\ket{\psi^+}_{B'C'}) = 0. 
\end{equation}
This superadditivity means that the bound entanglement is activated with the help of the auxiliary distillable entanglement, although A and D share no distillable entanglement for $\rho_s$ or $\ket{\psi^+}_{B'C'}$ alone. 
This protocol can also be regarded as an entanglement transfer from B-C to A-D.
In this context, it is interesting that two parties (B-C) can transfer the Bell state to the other two parties (A-D) despite being separated: $D_{AD|BC}(\rho_s) = 0$.  
In other words, entanglement can be transferred by the mediation of the undistillable, bound entanglement. 
This unique feature is quite different from two-stage entanglement swapping \cite{Goebel2008}, which needs distillable entanglement shared by senders and receivers to transfer the Bell states.

\begin{table}[b!]
\caption{Relationship between the combination of the results of the BSM of B-B' and C-C', and the state of AD. Each combination of the BSMs tells the state of A-D. }
\label{table}
\begin{center}\small
\def\arraystretch{1.8}
\begin{tabular}{lcccccc}\hline 
                     & &  & $\ket{\phi^+}_{BB'}$ & $\ket{\phi^-}_{BB'}$ & $\ket{\psi^+}_{BB'}$& $\ket{\psi^-}_{BB'}$ \\ \hline 
$\ket{\phi^+}_{CC'}$ & &  & $\ket{\psi^+}_{AD}$  & $\ket{\psi^-}_{AD}$  & $\ket{\phi^+}_{AD}$ & $\ket{\phi^-}_{AD} $ \\ 
$\ket{\phi^-}_{CC'}$ & &  & $\ket{\psi^-}_{AD}$  & $\ket{\psi^+}_{AD}$  & $\ket{\phi^-}_{AD}$ & $\ket{\phi^+}_{AD} $ \\ 
$\ket{\psi^+}_{CC'}$ & &  & $\ket{\phi^+}_{AD}$  & $\ket{\phi^-}_{AD}$  & $\ket{\psi^+}_{AD}$ & $\ket{\psi^-}_{AD} $ \\ 
$\ket{\psi^-}_{CC'}$ & &  & $\ket{\phi^-}_{AD}$  & $\ket{\phi^+}_{AD}$  & $\ket{\psi^-}_{AD}$ & $\ket{\psi^+}_{AD} $ \\ \hline
\end{tabular}
\end{center}
\vspace*{-4mm}
\end{table}

\begin{tiny}
\begin{figure}[t!]
   \includegraphics[width=\columnwidth, clip]{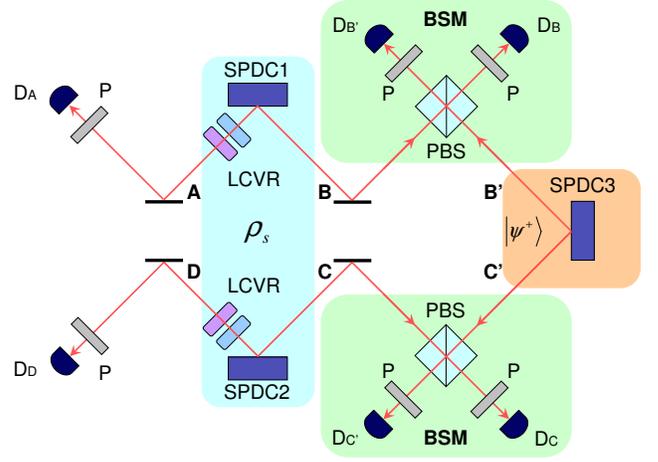} 
   \label{ }
   \caption{Scheme of the activation of the bound entanglement. 
Each source of spontaneous parametric down-conversion (SPDC) produces $\ket{\psi^+}$. 
The four-photon states emitted from SPDC1 and SPDC2 pass through liquid crystal variable retarders (LCVRs) to be transformed into the Smolin states.  
The polarization of two-photon states in mode A and D are analyzed on the condition that the pair of the Bell state, $\ket{\phi^+}_{BB'}\otimes\ket{\phi^+}_{CC'}$ is detected. P, a polarizer. 
 }
\end{figure}
\end{tiny}

Figure 2 illustrates the experimental scheme of our activation protocol. 
In our experiment, the physical qubits are polarized photons, having horizontal $\ket{H}$ and vertical $\ket{V}$ polarizations as the state bases. 
By using three sources of spontaneous parametric down-conversion (SPDC) \cite{Kwiat95}, we produced three $\ket{\psi^+}$ states simultaneously. 
The state $\rho(\psi^+) \equiv \ket{\psi^+}\bra{\psi^+}_{AB}\otimes \ket{\psi^+}\bra{\psi^+}_{CD}$ emitted from the SPDC1 and SPDC2 was transformed into the Smolin state by passing through the synchronized liquid-crystal variable retarders (LCVRs, see Supplementary Material).  
The state $\ket{\psi^+}_{B'C'}$ emitted from SPDC3 was used as the auxiliary Bell state for the activation protocol. 
A polarizing beam splitter (PBS) and a $\ket{+}_{B}\ket{+}_{B'}$ ($\ket{+}_{C}\ket{+}_{C'}$) coincidence event at modes B and B' (C and C') allow the projection onto the Bell state $\ket{\phi^+}_{BB'}$ ($\ket{\phi^+}_{CC'}$), 
where $\ket{+}_{i}= (\ket{H}_i+\ket{V}_i)/\sqrt{2}$  \cite{Goebel2008}. 
Given the simultaneous BSMs at B-B' and C-C' we post-selected the events of detecting $\ket{\phi^+}_{BB'} \otimes \ket{\phi^+}_{CC'}$ out of the 16 combinations (Table I). 
%
The result, i.e., the state after the activation, was expected to be $\ket{\psi^+}_{AD}$. 
To characterize the experimentally obtained Smolin state $\rho_s^{exp}$ and the state 
after the activation process  $\rho_{AD}$,
the maximum likelihood state tomography \cite{James2001} was performed (see Supplementary Material).


\begin{tiny}
\begin{figure}[t!]
   \includegraphics[width=0.7\columnwidth, clip]{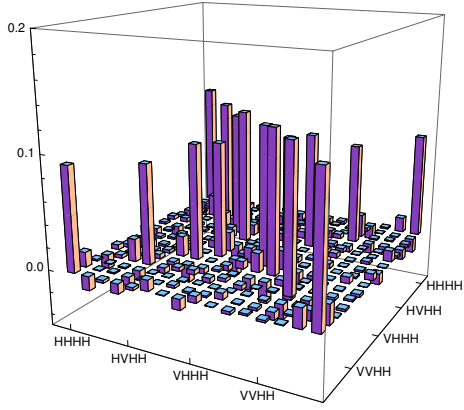} 
   \caption{ Real part of the measured Smolin state $\rho_s^{exp}$.    }
\end{figure}
\end{tiny}

Figure 3 shows the real part of the reconstructed density matrix of the Smolin state $\rho_s^{exp}$ we obtained.  
The fidelity to the ideal Smolin state $F_s$ = $ \left({\rm Tr} \sqrt{\sqrt{\rho_s} \rho_s^{exp} \sqrt{\rho_s} }\right)^2$ was calculated to be 82.2$\pm$0.2$\%$.
From $\rho_s^{exp}$, we evaluated the separability of the generated state across the bipartite cuts AB$|$CD, AC$|$BD, and AD$|$BC in terms of the logarithmic negativity (LN) \cite{Vidal2002}, which is an entanglement monotone quantifying the upper bound of the distillable entanglement under LOCC.
The LN of the density matrix $\rho$ composed of the two subsystems $i$ and $j$ is given by 
\begin{equation}
LN _{i|j}(\rho ) =\log_2 ||\rho^{T_{i}}||,
\end{equation}
where $\rho^{T_{i}}$ represents the partial transpose with respect to the subsystem $i$, and $||\rho^{T_{i}}||$ is the trace norm of $\rho^{T_{i}}$. 
The LN values thus obtained for the three bipartite cuts of $\rho_s^{exp}$ are presented in Table II, together with those of the Smolin state $\rho_s$ and the state $\rho(\psi^+)$. 
The state $\rho_s$ has zero negativity for all three bipartite cuts, while the state $\rho(\psi^+)$ has finite values, i.e., finite distillable entanglement originally from the two Bell states, for AC$|$BD and AD$|$BC cuts. 
For $\rho_s^{exp}$, the LN values are all close to zero, indicating that $\rho_s^{exp}$ has a very small amount, if any, of distillable entanglement. 

\begin{table}[b!]
\caption{Logarithmic negativities (LNs) for the two-two bipartite cuts. }
\label{table}
\begin{center}\small
\def\arraystretch{1.3}
\begin{tabular}{lccc}\hline 
                   & $\rho_s^{exp}$      & $\rho_s$ & $\rho(\psi^+)$ \\ \hline
$LN_{AB|CD}(\rho)$ & $0.076 \pm 0.012 $ & 0        &   0 \\ 
$LN_{AC|BD}(\rho)$ & $0.183 \pm 0.012 $ & 0        &   2 \\
$LN_{AD|BC}(\rho)$ & $0.178 \pm 0.012 $ & 0        &   2 \\\hline
\end{tabular}
\end{center}
\vspace*{-4mm}
\end{table}
To test other separabilities, we calculated the entanglement witness Tr$\{W \rho_s^{exp} \}$, which shows negative values for the non-separable states, and non-negative values for separable ones. 
The witness for our four-qubit states is $W = I^{\otimes 4}-\sigma_x^{\otimes 4}-\sigma_y^{\otimes 4}-\sigma_z^{\otimes 4}$ \cite{Amselem2009,Toth2005}. 
We obtained Tr$\{W \rho_s^{exp} \}= -1.30 \pm 0.02$, while the values for the ideal Smolin state and the state $\rho(\psi^i)$ were both -2. 
The negative witness value indicates that $\rho_s^{exp}$ has no separability in general. 
Taking account of the result that $\rho_s^{exp}$ has almost no distillable entanglement for the two-two bipartite cuts, $\rho_s^{exp}$ should have distillable entanglement in one-three bipartite cuts and/or tripartite cuts. 

\begin{tiny}
\begin{figure}[t!]
 \includegraphics[width=0.484\columnwidth, clip]{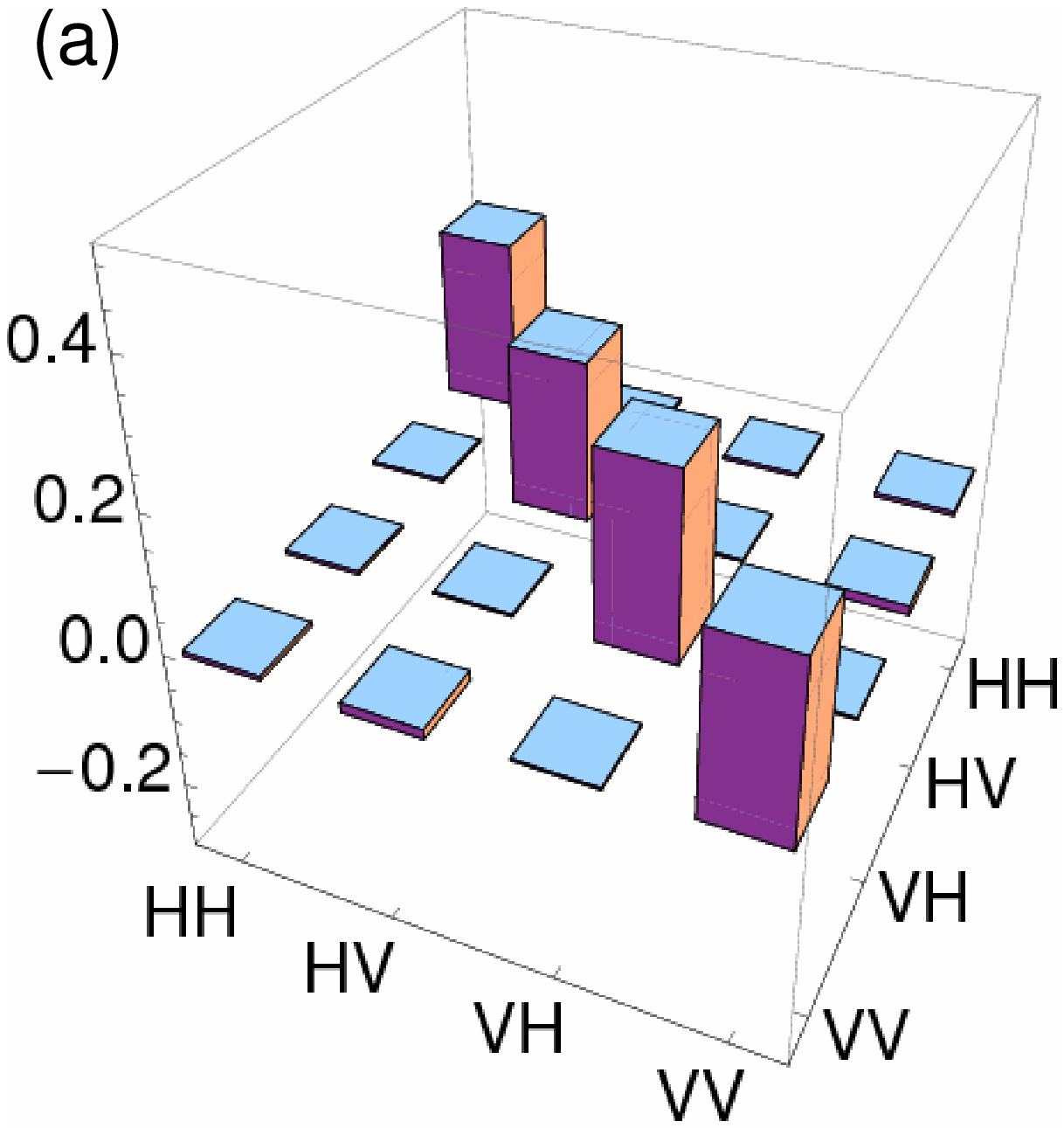} 
\includegraphics[width=0.49\columnwidth, clip]{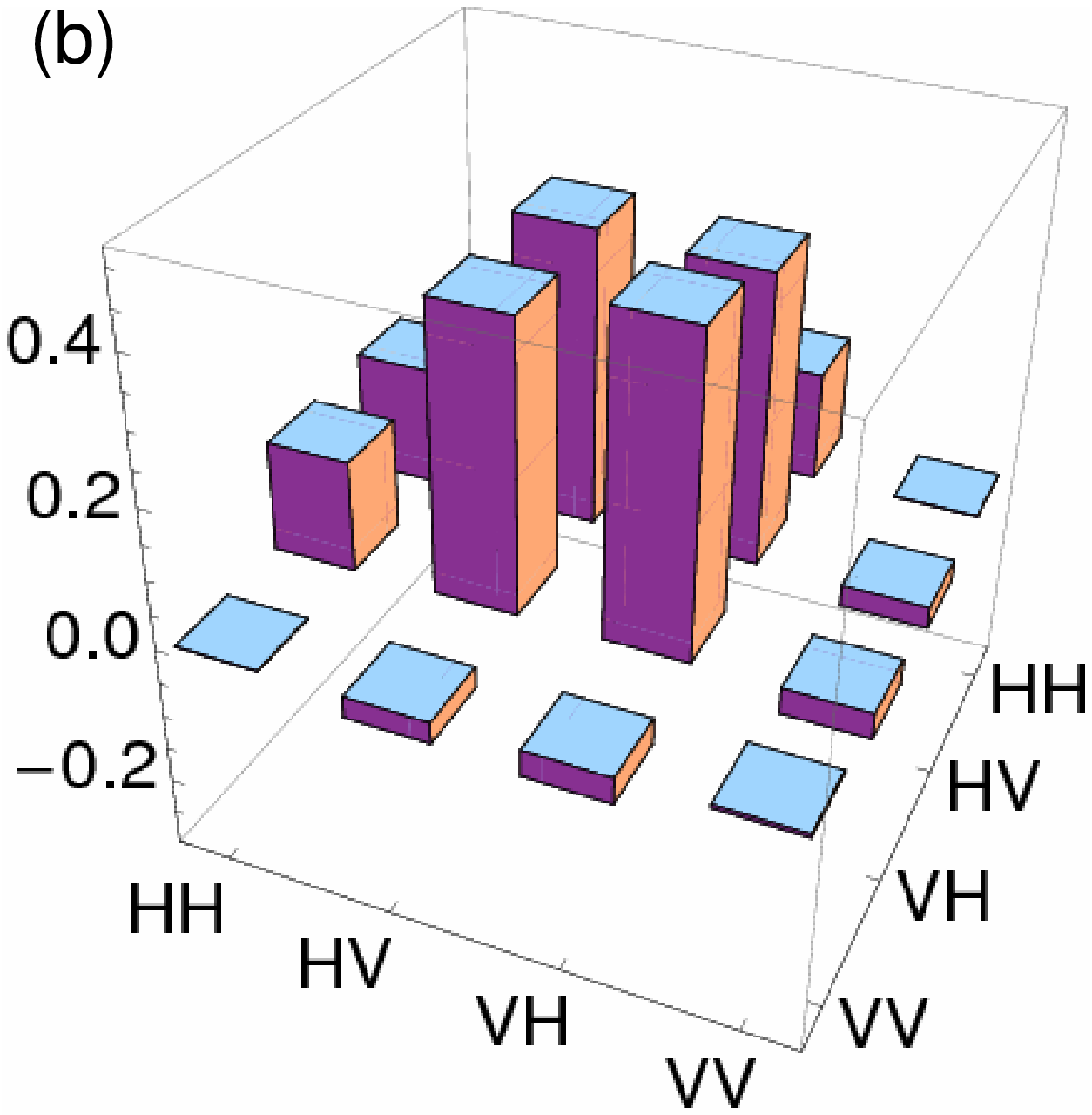}

     \label{ }
   \caption{Density matrices of the qubits A and D before (a) and after (b) the activation experiment. (a) Reduced-density matrix $ \rho_{sAD}^{exp}$ = Tr$_{BC} \rho_s^{exp}$. (b) The density matrix $\rho_{AD}$, triggered by the two BSMs at B-B' and C-C'.  }
\end{figure}
\end{tiny}
In the following we describe the results of our activation experiment. 
Figure 4 (a) shows the reduced density matrix $ \rho_{sAD}^{exp}$ = Tr$_{BC} \rho_s^{exp}$, i.e., the density matrix before the activation.  
We confirmed that $ \rho_{sAD}^{exp}$ gives no distillable entanglement to A and D: $ LN_{A|D}(\rho_{sAD}^{exp}) = 0$. 
Figure 4 (b) shows the density matrix after the activation $\rho_{AD}$, the reconstructed two-qubit density matrix in modes A and D triggered by the two BSMs at B-B' and C-C'. 
In the process of the state reconstruction we subtracted the accidental coincidences caused by higher-order emission of SPDC (see Supplementary Information). 
The fidelity $F_{AD} = \bra{\psi^+}\rho_{AD}\ket{\psi^+}_{AD} $ to the ideally activated state $\ket{\psi^+}_{AD}$ was 85$\pm 5$\%, which is larger than the classical limit of 50\%. 
The obtained LN was $LN_{A|D}(\rho_{AD}) = 0.83\pm 0.08$, indicating that we have gained a certain amount of entanglement via our activation process, whereas A and D initially share no distillable entanglement.

We quantified the increase of the distillable entanglement via our activation experiment. 
We evaluated the distillable entanglement before and after the activation by means of its lower and upper bounds as follows. 
The upper bound of $D_{A|D}(\rho_s^{exp})$, the distillable entanglement before the activation, was given by
\begin{align}
D_{A|D}(\rho_s^{exp}) \leq LN_{AB|CD}(\rho_s^{exp} ) = 0.076.   
\end{align}
The observed LN value after the activation process, $LN_{A|D}(\rho_{AD} ) = 0.83$, is larger than this value. 
However, since these are just the upper bounds of the distillable entanglement, we should examine the lower bound of $\rho_{AD}$ to confirm the increase of the distillable entanglement. 
To quantify the lower bound of $D_{A|D}(\rho)$, we used 
\begin{align}
D_H(\rho) \leq D_{A|D}(\rho ),
\end{align}
where $D_H(\rho)$ is the distillable entanglement via a certain distillation protocol, the so-called hashing method, known as the best method for Bell diagonal states of rank 2 \cite{Bennett1996}. 
$D_H(\rho)$ is given by, 
\begin{align}
D_H(\rho ) = 1+F\log_2(F)+(1-F)\log_2\frac{(1-F)}{3}, 
\end{align}
%
where $F$ is the maximum state-fidelity over the four Bell states $F$ = max$(\bra{\phi^i}\rho\ket{\phi^i})$. 
It is known that $D_H(\rho) > 0$ for $F > 0.8107$ \cite{Bennett1996}. 
The fidelity for $\rho_{AD}$ does satisfy this criterion and the value of $D_H(\rho_{AD})$ is calculated to be 0.15. 
The combination of Eq. (6) and (7) show a clear increase of the distillable entanglement via our activation experiment:  $D_{A|D}(\rho_s^{exp}) \leq 0.076 < 0.15 \leq D_{A|D}(\rho_{AD})$. 
%

%
In conclusion, we have experimentally demonstrated the activation of bound entanglement, unleashing the entanglement bound in the Smolin state by means of LOCC with the help of the auxiliary entanglement of the two-qubit Bell state. 
We reconstructed the density matrices of the states before and after the activation protocol by full state tomography. 
We observed the increase of distillable entanglement via the activation process, examining two inequalities that bind the values of the distillable entanglement. 
The gain of distillable entanglement clearly demonstrates the activation protocol in which the undistillable, bound entanglement in $\rho_s$ is essential. 
Our result will be fundamental for novel multipartite quantum-communication schemes, for example, quantum key secret-sharing, communication complexity reduction, and remote information concentration, in which general classes of entanglement, including bound entanglement, are important.  

This work was supported by a Grant-in-Aid for Creative Scientific Research (17GS1204) from the Japan Society for the Promotion of Science.

\clearpage
\section*{Supplementary Material for ``Experimental Activation of Bound Entanglement''}
\section{Photon sources}

To achieve high multi-photon count rates, we used the third harmonics (wavelength = 343~nm, average power = 350~mW, pulse duration = 250~fs) of the mode-locked Yb laser (Amplitude Systems, t-Pulse 500) as a pump source for SPDC. 
The pump source has higher pulse energy ($\sim$37 nJ) than those of typical multi-photon generation systems 
based on the second harmonics of Ti-Sapphire lasers (10$\sim$20 nJ) (e.g., Ref. [3]). 
Moreover, this pump source produces SPDC with a center wavelength of 686 nm, which is close to the spectral region where Si-avalanche photodiodes (Si-APD, Perkin Elmer SPCM-AQR-14) have maximum quantum efficiency ($\sim$65\%). 
The pump beam passes three Type-II beta barium borate (BBO) crystals (2-mm thickness) to produce three $\ket{\psi^+}$ states [24]. 
Each photon produced by the SPDC was passed through a band-pass filter (FWHM = 1~nm) for spectral filtering, and was led to a single-mode fiber for spatial filtering. 
The filtering processes make photons indistinguishable except for their polarizations. 
The average of n-fold coincidence rates $C_n$ were $C_1=2.4 \times 10^5$, $C_2=4.2\times 10^4$, $C_4=2.0 \times 10^2$ and $C_6=0.9$ sec$^{-1}$, respectively. 
The detection probability of photons per detection mode $\eta=C_2/C_1$ was 18\%. 
The probability of the SPDC per one pump pulse was calculated as $P = C_{6}/C_{4} \eta^2 = 0.15 $. 
The fidelities $F_{\psi^+}$ of the state emitted from each BBO crystal to the ideal $\ket{\psi^+}$ state were $F_{\psi^+} \geq 90\%$.

\section{Preparation of the Smolin states}

The four-qubit Smolin state shown in Eq. (1) is prepared as follows: 
$\rho(\psi^+) = \ket{\psi^+}\bra{\psi^+}_{AB}\otimes \ket{\psi^+}\bra{\psi^+}_{CD}$ is first emitted from two BBO crystals.  
The state $\rho(\psi^+)$ is randomly transformed to any $\rho(\phi^i)= \ket{\phi^i}\bra{\phi^i}_{AB}\otimes \ket{\phi^i}\bra{\phi^i}_{CD}$ by operating $\sigma_{\mu}^A \otimes \sigma_{\mu}^D$, where $\sigma_{\mu}^{\xi}$ is the one of the Pauli operations $\sigma_{\mu}= \{I, \sigma_x, \sigma_y, \sigma_z \}$ 
on the mode $\xi$. 
We implemented $\sigma_{\mu}^{\xi}$ using a pair of LCVRs, one of which could be set to $I$ or $\sigma_x$, and the other to $I$ or $\sigma_z$. 
Their combination can produce all Pauli operations. 
To guarantee the random and uniform statistical mixture of the four Bell states, i.e. the Smolin state, the retardation of each LCVR was set by pseudo-random number generators operating at a rate of 3 Hz.

\section{State Reconstruction}

We utilized the maximum likelihood method [25] for the reconstruction of the density matrices. 
To characterize the experimental Smolin state $\rho_s^{exp}$ we used the experimental setup shown in Fig. 2 in the main text, without the PBSs set at two BSM parts. 
The four-fold coincidences of the modes A, B, C and D were collected for 256 polarization projective measurements, each of which were collected for 4 min.

To reconstruct the state after the activation process, we recorded the six-fold coincidences, i.e., coincidence between A and D conditioned by the detection of $\ket{\phi^+}_{BB'}$ and $\ket{\phi^+}_{CC'}$, for 16 polarization projective measurements in the modes A and D. 
For each projection measurement, we integrated the signal for 4.5 h (72 h in total).  
The results for the H/V base projections, which correspond to the diagonal parts of the density matrix, are shown in Fig. 1. 
Based on several reports of experiments using the multi-photon SPDC state (e.g., \cite{Barz10}), we knew that higher-order emission events of SPDC were likely to affect our measurements as accidental coincidence events. 
Thus, we estimated the accidental coincidence rates originating from higher-order eight-photon states, which are represented by the blue portion in Fig. 1. 
The measured coincidence events (Fig. 1) for HH and VV projections were not expected for $\ket{\psi^+}_{AD}$ and obviously degraded the fidelity. 
One can see that these unexpected signals almost always originated from the higher-order emissions. 
By subtracting these estimated accidental coincidences, we obtained $\rho_{AD}$ (Fig. 4 (b) in the main text), containing only the six-photon states.  
These accidental coincidences from higher-order emissions can be reduced by the technical improvement of the setup, for instance, the use of photon-number resolving detectors such as transition edge sensors \cite{Fukuda09}. 
Thus, we adopted $\rho_{AD}$ as the state we obtained after the activation process, eliminating the higher-order accidental counts. 
In the case that raw coincidence data were used for the state reconstruction, the fidelity and the logarithmic negativity would be 64$\pm$4\% and $0.43 \pm 0.05$, respectively.

\begin{tiny}
\begin{figure}[!h]
   \includegraphics[width=1\columnwidth, clip]{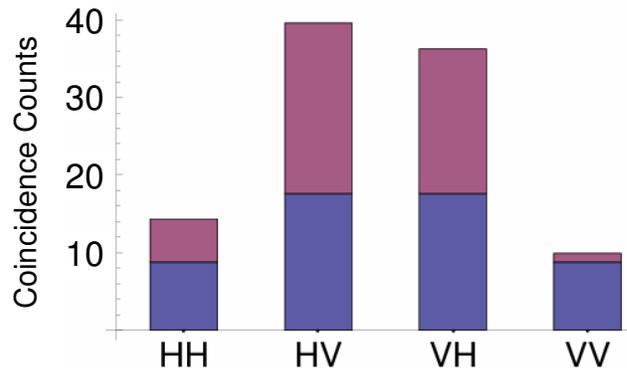} 
   \caption[10]{ Experimental six-fold coincidence distribution when the polarizations in A and D are analyzed in H/V basis (red). The rate of six-fold coincidences from higher-order emission of SPDC in each projection is estimated (blue). Almost all the measured undesirable coincidence events in HH and VV projections came from higher-order emissions. }
\end{figure}
\end{tiny}

\end{document}